%% file: paper.tex
\documentclass{llncs}

\usepackage{cite}
\usepackage{graphicx}
\usepackage{makeidx}  
\usepackage{todonotes}
\usepackage{subfigure}
\usepackage{tabularx}
\usepackage{amsmath}
\usepackage{tikz}
\usetikzlibrary{decorations.text,calc,arrows.meta, automata,arrows, positioning}
\usepackage{multirow}
\usepackage{hyperref}
\usepackage{listings}
\usepackage{url}
\usepackage{microtype}
\usepackage{xspace}
\usepackage{color}

\usetikzlibrary{calc}

\newcommand{\tool}[1]{{\sffamily{#1}}\xspace}

\begin{document}
\frontmatter          
\pagestyle{headings}  
\addtocmark{CounterExample Guided Neural Synthesis} 
%
%
%
\mainmatter              
\title{CounterExample Guided Neural Synthesis}
%
\titlerunning{CounterExample Guided Neural Synthesis}  

\author{%
Elizabeth Polgreen\inst{1} \and
Ralph Abboud\inst{2} \and
Daniel Kroening\inst{2}}

\authorrunning{Polgreen et al.}

\institute{University of California, Berkeley
\and
University of Oxford
}

\maketitle


\begin{abstract}
Program synthesis is the generation of a program from a specification. 
Correct synthesis is difficult, and methods that provide formal guarantees
suffer from scalability issues.  On the other hand, neural networks are able
to generate programs from examples quickly but are unable to guarantee that
the program they output actually meets the logical specification.  In this
work we combine neural networks with formal reasoning: using the latter to
convert a logical specification into a sequence of examples that guides the
neural network towards a correct solution, and to guarantee that any
solution returned satisfies the formal specification.  We apply our
technique to synthesising loop invariants and compare the performance to
existing solvers that use SMT and existing techniques that use neural
networks.  Our results show that the 
formal reasoning based guidance improves the performance of
the neural network substantially, nearly doubling the number of benchmarks
it can solve.


\end{abstract}

\section{Introduction}

Program synthesis is the task of automatically generating a program from a
given specification.  In its most general form, program synthesis is
complex, and scalable implementation is a challenge.  Existing methods
typically combine heuristic search of the space of programs together with an
SMT solver that asserts formally that the candidate program satisfies the
specification; a well-known instance is the CEGIS
loop~\cite{DBLP:journals/sttt/Solar-Lezama13}.  The key idea of
syntax-guided synthesis (SyGuS) is to ameliorate the scalability issue by
imposing restrictions on the search space by means of a syntactic template
provided by the user~\cite{DBLP:conf/fmcad/AlurBJMRSSSTU13}.

In parallel to these developments, machine learning is now known to be able
to solve the program synthesis problem effectively for a particular special
case: deep neural networks can be trained to produce string manipulation
programs given a set of examples of inputs and the desired program
outputs~\cite{devlin2017robustfill}.  While these approaches are fast, fully
specifying a non-trivial program using input-output examples only is
hard~\cite{DBLP:conf/icse/PelegSY18}, and thus, effective search for a
program that provably satisfies a given logical specification remains an
open problem.

In this paper we present two new algorithms that combine formal reasoning
using SAT/SMT solvers with a deep neural network in order to generate
programs that provably satisfy a full logical specification.
The first algorithm, \emph{Example-Guided Neural Synthesis}, comprises of
three steps: 1)~we use SMT to convert the given logical specification into a
set of input-output examples; the goal is to generate a set that captures as
much of the specification as possible; 2)~we use the neural network to
generate candidate programs from the set of input-output examples;
3)~finally, we use SMT to verify whether any of the candidate
programs returned by the network meets the specification.  If the network
does not return a correct program, we terminate with no answer.

We observe that the performance of the neural network is highly dependent on
the quality of the input-output examples provided.  We therefore hypothesise
that the neural network might benefit from a refined set of input-output
examples.  Thus, the second algorithm, \emph{CounterExample Guided Neural
Synthesis} (CEGNS), integrates the first algorithm into a CEGIS loop,
extracting counterexamples when the programs generated by the neural network
fail to meet the specification, and using these counterexamples to generate
input-output examples.  The loop iterates until the neural network generates
a correct program.  The loop is not guaranteed to terminate.

We evaluate our algorithms using a neural network trained to synthesise
predicates from input-output examples.  It is potentially possible to extend our
approach to functions returning bit vectors, or to other logics, though
further network designs need to be explored.  We evaluate our algorithms
on invariant synthesis benchmarks from the synthesis competition.  The key
finding is that the counterexample guidance significantly increases the
performance of the neural network: CEGNS nearly doubles the number of
benchmarks that the neural network can solve on its own.  Overall, CEGNS
performs comparably to a conventional CEGIS
implementation~\cite{DBLP:conf/cav/AbateDKKP18}, but solving a different
subset of the benchmarks.


The paper is structured as follows: Sections~\ref{sec:nn} and
\ref{sec:training} introduce the architecture of the neural networks we
build our algorithms around, and describes how we train them;
Section~\ref{sec:egns} and \ref{sec:cegns} present Example Guided Neural
Synthesis and CounterExample Guided Neural Synthesis;
Section~\ref{sec:experiments} describes the experimental setup, and
presents the experimental results; Section~\ref{sec:related} discusses
related work.

\section{Preliminaries}

\subsection{Program Synthesis}

Program synthesis tools solve a second-order existential logic problem that
can be formulated as follows:
$\exists{P} .\,  \forall{\vec{x}} .\,  \sigma(P,\vec{x})$
where $P$ ranges over functions, which are represented by means of programs,
$\vec{x}$ is a ground term corresponding to the input arguments for the target
program, and $\sigma$ is a quantifier-free formula that represents the
properties that must be satisfied by the program to meet the user's
specification.

%
%

There are many problems that can be expressed as synthesis problems of the
form above.  As an exemplar, we consider the problem of synthesising an
invariant for a given loop.  Let $\vec{x}$ denote a program state,
$I$~denote the predicate for the initial condition, $A$ an assertion, and
$T$ a transition relation.  Using the criteria of the invariant track in the
SyGuS competition, a predicate $P$ is an invariant if it is a model for the
following formula:
\begin{align*}
\exists P \,\,\forall \vec{x}, \vec{x'}.(I(\vec{x})
\Rightarrow P(\vec{x})) \,\wedge (P(\vec{x}) \wedge T(\vec{x}, \vec{x}')
\Rightarrow P(\vec{x}'))\, \wedge (P(\vec{x}) \Rightarrow A(\vec{x})). 
\end{align*}

%

\subsection{CounterExample Guided Inductive Synthesis}

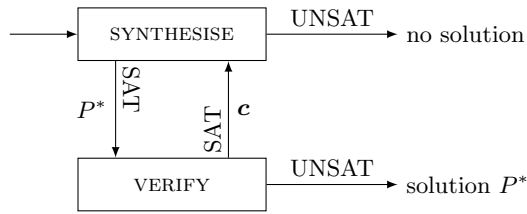
\begin{figure}[t]
\centering
\input{cegis.tikz}
\caption{CEGIS block diagram\label{fig:CEGIS_block}}
\end{figure}

CounterExample-Guided Inductive Synthesis (CEGIS), illustrated in
Fig.~\ref{fig:CEGIS_block}, is an iterative process consisting of two
phases: a synthesis phase and a verification phase.  Given the specification
$\sigma$ of the desired program, the inductive synthesis procedure generates
a candidate program $P^*$ that satisfies $\sigma(P^*,\vec{x})$ for a subset
$\vec{x}_\mathit{inputs}$ of all possible inputs.  The candidate program
$P^*$ is passed to the verification phase, which checks whether it satisfies
the specification $\sigma(P^*, \vec{x})$ for all possible inputs.  This is
done by checking whether $\neg\sigma(P^*, \vec{x})$ is unsatisfiable. 
If~so, then $\forall x.\sigma(P^*, \vec{x})$ is valid, and we have
successfully synthesised a solution and the algorithm terminates. 
Otherwise, the verifier extracts a counterexample~$\vec{c}$ from the
satisfying assignment, which is then added to the set of inputs passed to
the synthesiser, and the loop reiterates.

The verification phase is typically implemented using a SAT or a SMT solver. 
There is a wide range of options for the inductive synthesis component;
a~popular choice is heuristic enumeration in combination with an SMT
solver~\cite{DBLP:conf/tacas/AlurRU17}; another is synthesis using a SAT
solver~\cite{DBLP:journals/sttt/Solar-Lezama13}.  A well-known limitation of
enumeration-, SAT- and SMT-based techniques is their inability to synthesise
constants that are not given as part of the problem description, which can
potentially result in an inefficient enumeration of the search space.


\input{network.tex}

\section{Training the Neural Networks}
\label{sec:training}

\subsection{Sourcing the Training Data}
\label{sec:DataGen}

The performance of a neural network is dependent on the quality and quantity 
of the data it is trained on~\cite{dataprep}.
Neural networks typically require millions 
 of training examples~\cite{devlin2017robustfill,parisotto2016neuro}.  Given
we do not have millions of program synthesis benchmarks available, in order
to obtain a large training data set, we randomly generate predicates
constructed as syntactically correct combinations of bit-vector
instructions.  In an ideal world, we would randomly generate benchmarks in
the same format as the SyGuS benchmarks.  There are two reasons we elected
not to do this: first, randomly generating logical specifications that are
in some way similar to the problem we want to tackle is a hard problem; and
second, in order to train the network using supervised learning we need to
have both the logical specification and the solution to the logical
specification, and finding the solutions to invariant generation benchmarks
is hard.


Our neural network synthesises programs from I/O examples, where an I/O
example is an assignment to all input parameters of the program to be
synthesised, and a corresponding output that satisfies the specification. 
Our training data is a set of candidate programs, each accompanied by a
corresponding set of input-output examples.  We have experimentally
identified two features of ``bad training data'': 1)~multiple programs with
different syntax but equivalent semantics, which we term \emph{equivalent
mutants}\footnote{This term is borrowed from the literature on mutation
testing.}, and 2)~input-output examples that do not sufficiently
differentiate between programs.  To avoid bad training data, we supplement
our program generator with a set of syntactic rules and an SMT-based
procedure for both reducing redundancy and generating informative I/O
examples.

\subsubsection{Eliminating equivalent mutants}

We enforce several rules in our random program generator to reduce the
number of equivalent mutants generated.  As an example, shifting a bit
vector by any number greater than the bit vector width always produces the
same result as shifting the bit vector by the width.  Thus, we introduce a
rule such that no generated program contains any shift operation with the
second operand greater than the width of the first.


It would be impractical to check whether every program generated is
equivalent to any previous program generated using an SMT solver.  However,
we do use an SMT solver to identify a further cause of equivalent mutants:
An if-then-else statement that always returns one of its conditional outputs
is semantically equivalent to the same program without the if-then-else
statement, e.g., $(z=z) ?  x : y$ is equivalent to $x$.  We identify this
equivalence by recursively identifying leaf operators (i.e., operators
without nested \texttt{ite} statements) and checking their branch
satisfiability with an SMT solver.  If a branch is infeasible, we replace
the if-then-else statement with the other branch and apply the recursive
function again.  We continue until no leaf statements are redundant.

Further to this, we use an SMT solver to remove certain trivial programs,
namely programs that always return a single constant value, or always return
a single one of the input arguments.

\subsubsection{Input-Output Example Generation}
\label{ssec:IOGen}

The simplest way to generate input-output examples from a known target
program is to uniformly generate input values and to execute the program in
order to obtain the corresponding outputs.  This may not differentiate
between programs that do not represent a continuous function mapping the
input to output, i.e, they contain features that produce a discontinuous
output such as conditional statements.  We generate input-output examples
for each random program such that the examples are \emph{informative} with
respect to this program, i.e., the input-output covers as many conditional
branches in a program as possible, using a combination of randomly generated
inputs and SMT solving.  In order to obtain a distribution of inputs over
the input space when using an SMT solver, we use Z3 version 4.5.1~\cite{DBLP:conf/tacas/MouraB08} with the phase selection
set to random.

\subsection{Training the network}

To train the network, we randomly generate 32 million programs distributed
over programs of up to three input parameters, and of length up to give
operations, with up to one arbitrary constant, in addition to a zero and one
constant.


Of these 32 million programs, redundancy checks discard 2,613,768, leaving a
final training set of 29,386,232 programs.  For each program, we generate
ten input-output examples.  Training for the networks was done on an AWS
p3.xlarge instance with a Tesla Z100 GPU, and took 75 hours for the complex
network.  The network was trained once over the training data (i.e., for one
epoch) using the Adam optimiser~\cite{kingma2014adam} and masked
cross-entropy loss~\cite{mackay2003information}.


\section{Example Guided Neural Synthesis}
\label{sec:egns}

A limitation using neural networks to synthesise programs that meet logical
specifications is that neural networks are best suited to recognising
patterns in input-output examples rather than logical specifications, and,
given a specification and an input, it is not necessarily possible to
determine a single correct output for that input.  In Example Guided Neural
Synthesis, we use a combination of SMT solving and heuristics to generate
input-output examples that guide the network to the candidate solution, and
rely on the network's resilience to noise to overcome this limitation.  The
EGNS algorithm, illustrated in Fig.~\ref{fig:EGNS}, consists of three
components: An SMT-solver used to generate a sequence of input-output
examples from the invariant specification; a neural network as described in
Section~\ref{sec:network2}; and an SMT solver that checks each solution
generated by the network against the logical specification.

\begin{figure}[!htb]
    \centering
\begin{tikzpicture}[->,>=stealth' ,   state/.style={
           rectangle,
           rounded corners,
           draw=black, very thick,
           minimum height=2em,
           inner sep=4pt,
           text centered,
           text width=1.5cm
           }]
         
 \node[state] (IO) 
 {Example generator};
  \node[ left of=IO, node distance=1.9cm] (SPEC) 
 {spec}; 
 \node[state, right of =IO, node distance = 3.5cm] (SYNTH) 
 {Neural Network};
 \node[state, right of =SYNTH,  node distance=3.5cm] (VERIF) 
 {Verifier};
  \node[align=left,right of=VERIF, node distance=1.9cm] (OUT) 
 {final\\ program};
 \path 
 (SPEC) edge node [above]{} (IO)
 (IO) edge node [above, align=center]{input-output\\examples} (SYNTH)
 (SYNTH) edge node [above, align=center]{candidate\\programs} (VERIF)
 (VERIF) edge node [above]{} (OUT);

\end{tikzpicture}
    \caption{Example Guided Neural Synthesis}
    \label{fig:EGNS}
\end{figure}

\subsection{Generating Examples}
\label{sec:examples}

We use a combination of two approaches to generate input-output examples
from a logical specification: In general the approach we take is to randomly
generate possible input values across the full bitvector range and use an
SMT solver to determine a possible correct output from the program to be
synthesised.  If the number of possible correct outputs for a given input is
large, we may generate input-output examples that when combined preclude
viable candidate functions, and we rely on the statistical nature of the
neural network to address this problem.

In the particular case of invariants we are able to further reduce such
inconsistencies.  Recall that, given a predicate $P$ is an invariant iff
$\forall \vec{x}, \vec{x'}$
\begin{align*}
(I(\vec{x}) \Rightarrow P(\vec{x})) \,\wedge 
(P(\vec{x}) \wedge T(\vec{x}, \vec{x}') \Rightarrow P(\vec{x}'))\, \wedge 
(P(\vec{x}) \Rightarrow A(\vec{x}))
\end{align*}
For a given $x$, the result of $A(x)$, $T(x,x')$ and $I(x)$ are computable. 
We thus look for solutions to the following formula, where $P$ and $P'$ are
Boolean variables representing a correct output from $P(\vec{x})$ and
$P(\vec{x}')$ respectively:
\begin{align}
\exists P,P',\vec{x}' \, .\, (I(\vec{x}) \Rightarrow P) \,\wedge 
(P \wedge T(\vec{x}, \vec{x}') \Rightarrow P')\, \wedge 
(P \Rightarrow A(\vec{x}))
\label{eqn:output_ex}
\end{align}

In general it is not possible to determine the correct values of $P$ and
$P'$ (Figure~\ref{fig:invariant}).  We know $P(\vec{x})$ must be true if
$\vec{x}$ is part of the initial conditions and does not violate the
property, and that $P(\vec{x})$ must be false if $\vec{x}$ violates the
property, and that $P(\vec{x})'$ must be false if $\vec{x}'$ violates the
property.  There is a region of the state space where a valid $P$ should be
true because the state is reachable but we cannot determine that from a
single input value $\vec{x}$, and a region of the state space where a
satisfying invariant could be either true or false because the states are
unreachable but still satisfy the property.
%
%
We test three heuristics: in the first we only generate example pairs that
are definitively true or false; in the second we over-approximate the
invariant, and assume that it returns true for every case where $\vec{x}$
and $\vec{x}'$ satisfy $A(\vec{x})$; and in the third we under-approximate
the invariant and assume it returns false for every case where $\vec{x}$ and
$\vec{x}'$ do not satisfy $I(\vec{x})$.  Note that due to the statistical
nature of the neural network, the use of incorrect input-output examples
does not necessarily prevent the neural network from giving a correct
answer.  We find that the best results are obtained from a combination of
all three heuristics, i.e., when we cannot determine the output, we randomly
choose true or false.

If there exists an $\vec{x}$ that is both in the initial conditions and does
not satisfy the property, Formula~\ref{eqn:output_ex} is not satisfiable and
there no possible invariant.  There are no benchmarks in the set that meet
these criteria.

\begin{figure} \centering
\begin{tikzpicture}
\coordinate (O) at (0,0);
\draw (O) circle [x radius=3.5, y radius=1.7];
\draw (O) circle [x radius=2.5, y radius=1.1];
\draw (O) circle [x radius=1.0, y radius=0.5];

\node at (0,0.2) {$I(\vec{x})$};
\node at (0,-0.1) {\emph{known, true}};
\node at (0,0.7) {reachable};
\node at (0,-0.7) {\emph{unknown, true}};
\node at (0,-1.3) {\emph{unknown, don't care}};
\node at (0,1.4) {$A(\vec{x})$};
\node at (0,-1.9) {\emph{known, false}};

\end{tikzpicture}
\caption{Invariant valuation over a 2 dimensional $\vec{x}$}
\label{fig:invariant}
\end{figure}

A limitation of randomly generating inputs across the full range of the bit
vectors is that we may never hit an input for which the invariant is known
to definitely be true or false.  In the second approach, we guarantee that
we generate inputs for which the invariant is both true and false by using a
SAT solver to find values for $\vec{x}$ while asserting that either the
initial conditions are true, or that the property is false and the
transition function is true.

\section{CounterExample Guided Neural Synthesis}
\label{sec:cegns}

A natural extension to Example Guided Neural Synthesis is CounterExample
Guided Neural Synthesis, where the network is integrated into an iterative
process similar to CEGIS and the next set of examples are chosen based on
the previous set of incorrect programs that were generated by the network,
in order to try to guide the network to a program that meets the
specification.  The use of a counterexample-guided loop allows us to fix any
incorrect outputs gradually.

%

An overview of this architecture is given in Fig.~\ref{fig:CEGISDeepSynth}. 
The loop is initialised by generating a set of random input-output examples
as described in Section~\ref{sec:examples}.  We over-approximate the
invariant when generating the output examples.  This approximation is
corrected by counterexamples in subsequent iterations, inspired by
IC3~\cite{DBLP:journals/fac/BradleyM08}.

\begin{figure}[!htb]
    \centering
\begin{tikzpicture}[->,>=stealth' ,   state/.style={
           rectangle,
           rounded corners,
           draw=black, very thick,
           minimum height=2em,
           inner sep=4pt,
           text centered,
           text width=1.5cm
           }]
         
 \node[state] (IO) 
 {Example generator};
  \node[left of=IO, node distance=1.9cm] (SPEC) 
 {spec}; 
 \node[state, right of =IO, node distance = 3.5cm] (SYNTH) 
 {Neural Network};
 \node[state, right of =SYNTH,  node distance=3.5cm] (VERIF) 
 {Verifier};
  \node[align=left,right of=VERIF, node distance=1.9cm] (OUT) 
 {final \\ program};
 \path 
 (SPEC) edge node [above]{} (IO)
 (IO) edge node [above, align=center]{input-output\\examples}(SYNTH)
 (SYNTH) edge node [above, align=center]{candidate\\programs} (VERIF)
 (VERIF) edge node [above]{} (OUT)
 (VERIF) edge [bend left] node [below, align=center] {counterexample\\input-output} (IO);
\end{tikzpicture}
    \caption{CounterExample Guided Neural Synthesis}
    \label{fig:CEGISDeepSynth}
\end{figure}

\subsection{CounterExample choice}

It is known that counterexample choice is key for performance in
CEGIS~\cite{DBLP:journals/corr/JhaS14a}.  This is especially true with
neural network synthesisers, as these generalise best when deployed on
inputs that are similar to its training inputs.  In this paper, the network
trains for synthesis using very well-formed input-output sets which, by
construction, cover all execution branches and are restricted to the
program's working input range, and we endeavour to replicate these
informative input-output sets as far as possible inside the CEGNS loop.

\subsubsection{Probable candidate programs}

The first heuristic we introduce for counterexample selection takes
advantage of the complex neural network's ability to produce not just one
candidate program, but the $k$ most probable candidate programs.  Given $k$
candidate programs and a set of input-output examples, we introduce a
pre-verification step which checks each candidate starting with the most
probable candidate, against the set of input-output examples.  It continues
checking candidates until it has found $n$ candidates that satisfy all the
input-output examples, where $n$ is a heuristic parameter given by the user,
or until it has checked the $k^{th}$ program.  It returns these programs that
satisfied all input-output examples to the verifier.  In the event that no
candidates satisfy all input-output examples, it return the
candidate which satisfies the most examples.

\subsubsection{Distribution of counterexamples} 

In order to obtain a distribution of counterexamples over the possible
values that is as similar as possible to the training data, we use Z3 as a
verifier with the same random phase-selection used for generating the
training data.  Though this does not guarantee similarity with training
input-output, owing to Z3 solving a different and more constrained problem
when generating counterexamples, it does increase the variance in
counterexamples.

\subsubsection{Generation of output examples}

Each counterexample consists of values assigned to the inputs $\vec{x}$ for
which the invariant returned the wrong answer.  To convert this to an
input-output example we take the output produced by the counterexample
inputs and negate it.  In the case where we receive a counterexample due to states that
violate the inductiveness of the candidate invariant (i.e., violate
$P(\vec{x}) \wedge T(\vec{x},\vec{x'}) \Rightarrow P(\vec{x'})$),
conventional model checking algorithms must make a choice whether to remove
$x$ from the invariant or to add $x'$. Typically they act in a monotonic way, doing either one
or the other.  Since we initialise the loop with an over-approximation of the
invariant, we opt for providing the example that guides the network towards
removing $x$ from the invariant.  However, due to the statistical nature of
the neural network, and the fact that the sequence of candidate invariants
it generates will not be monotonic, this is not guaranteed to be the correct
answer.  As a measure to compensate for potentially generating input-output
examples that are not compatible with a single invariant, we maintain a
finite number of counterexamples, discarding older counterexamples as new
ones are obtained.  We~also discard duplicate counterexamples.

\section{Experimental Results}
\label{sec:experiments}

\subsection{Experimental Setup}
\label{sec:Setup}

We use a 12-core 2.40\,GHz Intel Xeon E5-2440 with 96\,GB of RAM running
Linux.  We implement the counterexample generation, the example generation
and the program verification in C++ and the neural network in Python.  We
evaluate Example Guided Neural Synthesis and CounterExample Guided Neural
Synthesis using a two sets of benchmarks: 1)~problems from the loop
invariant category taken from the Syntax Guided Synthesis Competition where
we have replaced unbounded integer types by bit-vector types; 2)~benchmarks
that correspond to safety invariants for C
programs~\cite{DBLP:conf/fm/DavidKKL16}.  We apply a timeout of 600s. Our code, benchmarks and the 
scripts in order to run the experiments are available to 
download\footnote{\url{https://drive.google.com/open?id=1VkCyy7Sbipymn353R4g0BquXORjjs4dL}}.

We compare with \tool{CVC4}~\cite{DBLP:conf/cav/BarrettCDHJKRT11} version
1.7-prerelease [git master dd9246f3] (CVC4 1.5 was the clear winner of the
general track of the 2018 competition), our implementation of a standard
CEGIS loop~\cite{DBLP:conf/fm/DavidKKL16} with an SMT solver as the verifier
and CEGIS(T)~\cite{DBLP:conf/cav/AbateDKKP18}.

We compare the two best configurations of CEGNS and EGNS and the
solver-based tools.  There were two tools that we would have liked to
compare to but were unable to; the first is \textsf{LoopInvGen}, which
narrowly beats \tool{CVC4} in the invariant track of the SyGuS
competition~\cite{code2inv}. 
\tool{LoopInvGen}~\cite{DBLP:conf/pldi/PadhiSM16} is limited to arithmetic
over unbounded integers; consequently, we ran it on equivalent benchmarks
translated to use linear integer arithmetic, of which it solved 67 in
$\sim$10\,s per benchmark.  However, owing to the semantic gap, some of the
benchmarks have different solutions, and it is unclear whether the
performance observed is achievable on bit vectors.  The second tool is
\tool{Code2Inv}~\cite{code2inv}, a tool based on neural networks for
synthesising loop invariants for C code.  Even though \tool{Code2Inv} takes
C files as input, it treats C integers as unbounded integers. 
Unfortunately, it was unable to parse our benchmarks.

\subsection{Results}

We empirically sample across the following parameters controlling how counterexamples
are generated:
\begin{enumerate}
\item $\mathit{beam}$: the number of programs produced by the network at once, e.g., a
beam of size of $k$ means that the network generates the $k$ most likely
programs.
\item $\mathit{progs}$: the maximum number of programs returned by the network to the
verifier, as explained in Section~\ref{sec:cegns}.
\end{enumerate}
We find that storing the maximum number of programs possible after each
network call allows us to solve a few more benchmarks than only storing the
first one that satisfies all the input-output examples, although it is
slower.  A beam size of $100$ solves more benchmarks than a beam size
of~$10$.  We opt for maximising the number of benchmarks solved, in our
comparison in our comparison with other solvers shown in
Table~\ref{tab:results}, and use a beam size of $100$, and maximally store
up to $100$ programs per network call.

%
%

\begin{table}[!htb]
	\begin{center}
	\caption{Comparison of CEGNS with the state of the art}
	\begin{tabular}{l@{\quad}r@{\quad}r@{\quad}r@{\quad}r@{\quad}r@{\quad}r}
		\hline
		 & CEGNS & EGNS & CEGIS & CEGIS(T) & CVC4 \\
		\hline 
		Benchmarks solved &  21   & 12    & 23   & 26   & 53 \\
        Average time (s)  &  46.3 \,s  & 13.7\,s & 1.5\,s & 74.3\,s & 5.5\,s \\
		\hline
	 \label{tab:results}
	\end{tabular}
    \end{center}
\end{table}

Example Guided Neural Synthesis solves only 12 benchmarks with beam size
$100$ although the runtime is quick.  CEGNS solves nearly twice as many
benchmarks as EGNS, demonstrating that the counterexample guidance improves
the performance of the neural network, and given a more performant network
CEGNS could be a powerful tool.  CEGNS solves a similar number of benchmarks
to our traditional solver-based CEGIS implementation, and solves five
benchmarks that CEGIS fails to solve, and two that CEGIS(T) fails to solve.

CVC4's performance based on single invocation in formidable, in terms of
both speed and benchmarks solved. 
However, in cases where our network and CVC4 both find a solution, the
solution given by the network is typically shorter and more human readable. 
This is likely due to the style of the training data, and is synthesis
biased towards human readable answers is an area worth exploring.

\subsection{Threats to validity}

\textit{Benchmark selection: } We report an assessment of our approach on a
diverse selection of benchmarks.  Nevertheless, the set of benchmarks is
limited within the scope of this paper, and the performance may not
generalise to other problems. \\
\textit{Optimality of neural network architecture: } We evaluate CEGNS and
EGNS on an empirically chosen neural network architecture; other
architectures may be more performant.  The architecture was chosen on the
basis of existing literature.
\textit{Choice of theories: } CEGNS requires a neural network that is
trained for the relevant theory, e.g., linear integer arithmetic, bit
vectors etc.; our evaluation is limited to the theory of bit vectors, and
results for other theories may be worse.

\section{Related Work}
\label{sec:related}

Research applying neural networks to program synthesis typically focuses on
synthesising programs from I/O examples.
A main approach for this is \emph{supervised learning}.

\par For instance, Parisotto et al.~introduce the \tool{NSPS}
(Neuro-Symbolic Program Synthesis) system~\cite{parisotto2016neuro}, which
synthesise string manipulation programs from I/O examples using a
Recursive-Reverse-Recursive Neural Network that synthesises programs by
incrementally expanding partial programs.  Neural
GPUs~\cite{DBLP:journals/corr/KaiserS15} are capable of learning to perform
discrete algorithmic tasks such as reversing a sequence but the program
itself cannot be extracted.
%
%
\tool{RobustFill} also synthesises string manipulation
program~\cite{devlin2017robustfill}.  The architecture of the neural network
of \tool{RobustFill} is similar to that we use in CEGNS.  Supervised
learning methods require a large set of programs and I/O examples to train,
which can be difficult to generate or acquire, but are very fast to query
once trained.

Reinforcement learning used to train neural networks for program
synthesis~\cite{abolafia2018neural,bunel2018leveraging} does not require
I/O examples to generate programs, and uses a solution checker to generate
the reward needed for reinforcement learning; however, the programs
synthesised are typically in restrictive languages with 10 or fewer
operators, so it is not clear whether this approach would scale to
synthesising invariants for bitvector programs.

Neural networks have also been used to supplement, rather than replace,
conventional program synthesis techniques.  For example, Balog et
al.~\cite{balog2016deepcoder} develop \tool{DeepCoder}, a program synthesis
engine that uses a neural network to guide and augment conventional search
techniques.

\tool{Code2Inv} is a neural network based tool specifically for synthesising
loop invariants for C programs~\cite{code2inv}.  The network is designed to mimic a
human synthesising an invariant by working in a compositional fashion.  The
reported results are promising, but the tool will not work for loops in C
programs where overflow semantics are critical to the safety of the loop,
and in testing, we were unable to run the tool on any of our benchmarks.

SMT solvers are frequently used as oracles in formal program synthesis,
including in the CEGIS algorithm first presented for program
sketching~\cite{DBLP:conf/asplos/Solar-LezamaTBSS06,DBLP:journals/sttt/Solar-Lezama13}. 
Component-based approaches to program
synthesis~\cite{DBLP:conf/pldi/PerelmanGGP14,DBLP:conf/pldi/GulwaniJTV11,DBLP:conf/pldi/GulwaniKT11,DBLP:conf/pldi/FengMGDC17,DBLP:conf/popl/FengM0DR17,DBLP:conf/ndss/FengBMDA17,DBLP:conf/cav/AlbarghouthiGK13}
are of interest to us as we feel neural networks would be equally capable of
learning how to assemble programs from a database of components rather than
from a set of instructions.  Component based synthesis typically makes use
of techniques from counterexample-guided
synthesis~\cite{DBLP:conf/pldi/GulwaniJTV11} to type-directed search with
lightweight SMT-based deduction and partial
evaluation~\cite{DBLP:conf/pldi/FengMGDC17} and
Petri nets~\cite{DBLP:conf/popl/FengM0DR17}.
 
There exist many algorithms specifically for invariant generation, and,
although our algorithm has potential to be extended to more general program
synthesis, it is worth mentioning the similarities.  Our counterexample
generation resembles IC3~\cite{DBLP:journals/fac/BradleyM08} in the way it
refines invariants.  \tool{LoopInvGen}~\cite{DBLP:conf/pldi/PadhiSM16} uses
program synthesis based techniques to synthesise loop invariants, and is
similarly data-driven.  constraints.

\section{Conclusions}
\label{sec:conclusions}

We have introduced an Example Guided Neural Synthesis algorithm and a
CounterExample Guided Neural Synthesis Algorithm.  Both allow application of
neural synthesis to logical specifications.  CEGNS is a promising step towards
guaranteed correct neural-network based program synthesis, and performs comparably to a
purely SMT based CEGIS implementation, despite having a poorly performing
network as a core component. We look forward to exploring use of counterexample guidance
with more sophisticated network designs.
CEGNS has potential to be extended to more
general program synthesis problems, particularly if used in combination with
a technique such as CEGIS(T), where the counterexamples returned provide
more information than a single counterexample value.

\bibliography{Bibliography,paper} 
\end{document}

%% file: cegis.tikz
\begin{tikzpicture}[>=latex,x=3cm,y=2cm]
\node[rectangle,draw,minimum height=0.7cm,minimum width=2.5cm] at (1,1) (s) {\textsc{synthesise}};
\node[rectangle,draw,minimum height=0.7cm,minimum width=2.5cm] at (1,0) (v) {\textsc{verify}};
\node[anchor=west] at (2,1) (ns) {no solution};
\node[anchor=west] at (2,0) (so) {solution $P^*$};

\path[->] ($(s.west)+(-0.3,0)$) edge ($(s.west)+(0,0)$);
\path[->] (s) edge[above] node {UNSAT} (ns);
\path[->] (v) edge[above] node {UNSAT} (so);
\path[->] ($(v.north)+(0.25,0)$) edge[right] node {$\vec{c}$} node[left,near start,rotate=90,anchor=base,yshift=0.1cm] {~SAT} ($(s.south)+(0.25,0)$);
\path[->] ($(s.south)+(-0.25,0)$) edge[left] node {$P^*$} node[right,near start,rotate=270,anchor=base,yshift=0.1cm] {~SAT}($(v.north)+(-0.25,0)$);

\end{tikzpicture}

%% file: network.tex
\section{Neural Networks for Synthesising Programs}
\label{sec:nn}

A neural network consists of a set of neurons that aggregate input values
using a weighted linear combination and a non-linear activation function to
return a single output.  The output of a neuron with $k$ inputs is
\[ f(\Sigma_{i=1}^{k}(w_i\cdot x_i) + b) \]
where $w_i$ denotes the weight for the $i^\mathrm{th}$ input, $x_i$~denotes
the value of the $i^{\mathrm{th}}$ input, $b$~denotes the bias and $f$ is
the non-linear activation function.

The input to the neural network in our algorithms is a set of multiple I/O
\emph{examples}.  An I/O example is an assignment to the input parameters of
the function to be synthesised and a corresponding correct output.  The
input parameters are bitvectors of width 32, and we feed the network both
the binary representation of the bitvector and the decimal representation of
the value normalised to be between $-1$ and $1$.  The output assignment in
each example is a Boolean value.  Discrete values are represented using an
embedding that maps them to real numbers.  The use of such embedding
functions is standard.

The output of the neural network is a candidate program, represented as a
sequence of program \emph{tokens}.  The network has a vocabulary of 50
possible program tokens, 16 of which are used to represent constants, and
the remaining 34 represent program instructions, parentheses, and indicators
for ``start of program'' and ``end of program''.

We empirically explored a variety of neural network architectures. 
We~describe a simple network to illustrate our ideas, and a more complex
network with similar architecture to that used in neural program synthesis
for string manipulation programs~\cite{devlin2017robustfill}.  The simple
network fails to solve any but the most trivial benchmarks (e.g., where true
is a sufficiently strong invariant), and so our experimental work uses
the complex network.

\subsection{Network 1 -- A Simple Network}

\begin{figure} \centering
\hspace*{-0.3cm}\input{simple_network.tikz}
\caption{Network 1: the simple network} 
\label{fig:simple_net}
\end{figure}

Network 1 is a simple \emph{feed-forward} neural network.  In a feed-forward
network, neurons are organised into layers, where all neurons in a layer
have access to the same inputs; however, each neuron has its own set of
weights.  Information in a feed-forward network flows forward through these
layers, such that the output of a layer is the input for the next.  No
recurrence is used, and thus no neuron is used more than once in the
computation of the network's output.  Owing to this structure, feed-forward
architectures can only processes data sequences of bounded length.  Data
given to feed-forward networks must be preprocessed to fixed length.

We set up Network~1 such that it synthesises functions with up to three
input parameters: any missing input parameters are filled with padding data. 
The padding data is a learnable variable of the network: it can thus learn a
value that best encodes the non-existence of the corresponding parameter.
Network 1, given in Fig.~\ref{fig:simple_net}, consists of 
\begin{enumerate}
\item a feed-forward network for encoding;
\item a max-pooling operation aggregating the encodings for each input-output
example;
\item a feed-forward network for decoding.
\end{enumerate}
All feed-forward network layer neurons use a ReLU activation function, which
is $\max\{0,x\}$.  This activation function is
commonplace~\cite{Goodfellow-et-al-2016}.  The encoder processes an I/O
example as follows: it concatenates the inputs and the output into a single
vector.  This vector is fed into a feed-forward network with two neuron
layers; a first is a hidden layer and the second is the output layer.
The output of this network is a vector of floating-point values called the
\emph{encoding}, which summarises the I/O example the network is given.

To process multiple I/O examples, the encoder network is applied separately
to each example, and so produces an encoding for each example.  These
encodings are aggregated using the \emph{max-pooling operation}, which
combines the encodings by taking the maximum value from each element of the
vector.  The output $O$ of the max pooling operation applied to $k$
encodings is computed as follows:
\[O_{[j]} = \max_{1\leq i \leq k} \mathit{Enc}_i[j]\]
where $\mathit{Enc}_i[j]$ is the $j^\mathrm{th}$ dimension of the
$i^\mathrm{th}$ encoding vector.  The max pooling operator is chosen based
on its successful use in~\cite{devlin2017robustfill, parisotto2016neuro}. 
Using pooling, the network can process an arbitrary number of I/O examples
by computing the encoding using the encoder network and consolidating all
encodings via the pooling operation.

The decoder is another feed-forward network with two neuron layers.  The
decoder accepts the aggregated encoding as input, and maps this to a
sequence of program tokens.  The final layer of this network contains 5000
neurons, which can represent programs of up to 100 tokens long, using our
vocabulary of 50 tokens.  Each block of 50 neurons generates a single token. 
We use the softmax function to convert these floating-point outputs into a
discrete probability distribution over the vocabulary for each token.  The
final program is computed as the most likely token obtained from each block
of 50 neurons, with the exception of the first token, which yields the ``end
of program'' symbol.

This neural network architecture can scale to an arbitrary number of I/O
examples, while also being very fast to run and train due to its
feed-forward structure.  However, it lacks representational power.  For
instance, the decoder network does not explicitly enforce a dependence
between distinct program tokens dictated by its output, whereas this
dependency is evident in programs, e.g., the token ``bvgt'' is often
followed by a constant.  Furthermore, the network does not explicitly
account for the order of input parameters, which is critical for
non-commutative program synthesis.


\subsection{Network 2 -- A Complex Network}
\label{sec:network2}

Recurrent Neural Networks(RNNs) are designed to recognise patterns in
sequences of data, and have a temporal dimension.  Information is cycled in
a loop through the network, and the decision a recurrent network reached at
time step $t-1$ affects the decision it will reach at the next time step. 
This memory allows the network to find correlations between events that are
separated by many moments in time, and means that they can consider
sequences of variable length.  This ability to consider a sequence of
arbitrary length enables us to consider programs with arbitrary numbers of
input parameters.  Network 2 is based on a recurrent neural network.

Long-Short Term Memory (LSTM) cells~\cite{hochreiter1997long} allow RNNs to
learn patterns from sequential data more efficiently, in particular with
respect to long-term dependencies~\cite{bengio1994learning}.  LSTM cells
compute a \emph{state} representation at every time step of an input
sequence which describes the information it has processed thus far.  This
state is represented using two vectors: the \emph{cell state} and the
\emph{hidden state}.  The cell state serves as memory to retain knowledge
over time, while the hidden state is primarily used to produce the cell's
output.  LSTM cell states are updated using a combination of linear
operations (e.g., multiplication and summation), which are easily
differentiable and whose gradient does not decay over time, unlike most
non-linear activations.  This helps the network retain information over a
larger number of time steps.  An LSTM can be conditioned on the output of a
previous cell by using the final cell state and hidden state of the previous
cell as the current cell's initial hidden state and cell state.  Network~2,
shown in Fig.~\ref{fig:DSNN}, consists of, for each input-output example:

\begin{figure}
\centering
\includegraphics[width=12cm]{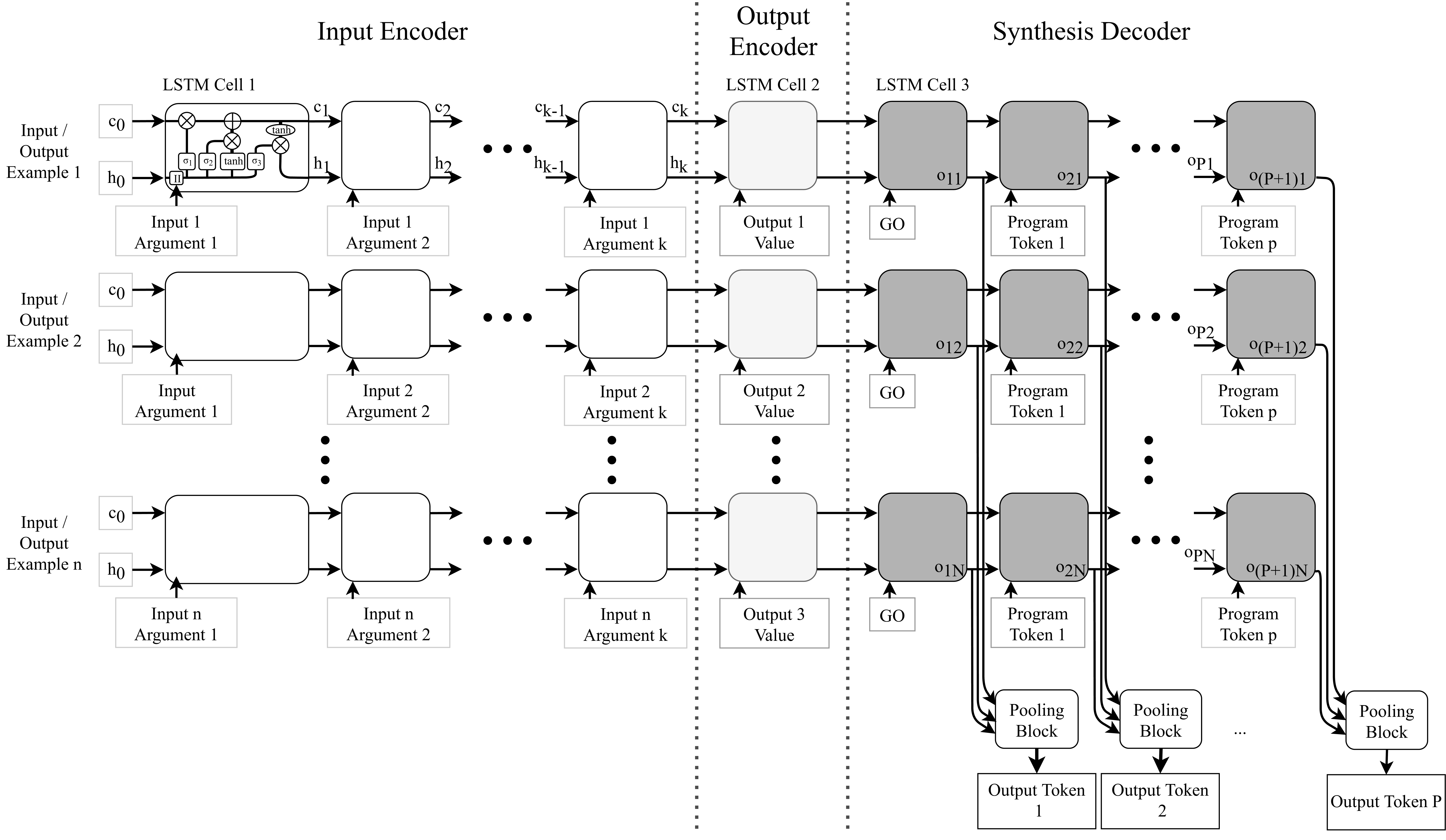}
\caption{Network 2: the complex network}
\label{fig:DSNN}
\end{figure}

\begin{enumerate} 
\item an \textbf{input encoder} LSTM recurrent network, which receives the
input parameter values from the input-output pair, fed in sequentially;
\item an \textbf{output encoder} LSTM cell, which receives the Boolean
output embedding value from the input-output pair and is conditioned on the
final cell and hidden states of the input encoder network;
\item a \textbf{decoder} LSTM recurrent network, which is trained on the
\textbf{target program} and conditioned on the final output encoder state.
\end{enumerate}
This is a typical architecture for sequence-to-sequence
networks~\cite{DBLP:journals/corr/SutskeverVL14}.

The final program is obtained by aggregating the outputs of the decoders
using pooling.  We then use either greedy decoding on the output of the
pooling, similar to Network 1, or Beam Search, which is a compromise between
performing a time-consuming complete search to find the optimal output
sequence and performing a speedy but sub-optimal greedy decoding.  Beam
search returns $k$ output sequences, where $k$ is referred to as the
\emph{beam size}.  Beam search computes the set of $k$ likely candidate
sequences at each time step until all $k$ candidates have reached an
end-of-sequence token, or are longer than a pre-defined limit.
 
Though pooling the outputs at this later point incurs extra computational
cost owing to the additional decoders, this approach delivers better results
in practice~\cite{devlin2017robustfill}.  Furthermore, the architecture
lends itself to extensions, in particular with regards
to incorporating \emph{attention}~\cite{chorowski2015attention}.  We choose a
recurrent synthesis decoder to overcome the limitations of MLPs, namely the
lack of explicit time-dependence between the decoders' tokens and the
structural upper bound on output length, albeit at the expense of
computational efficiency during training.

To train Network 2, the output of the decoder is compared with the target
sequence using the cross-entropy loss function, which is computed as
follows:
\[L = - \Sigma_{i=1}^{k} y_i \cdot \log(o_i)\]
where $o$ and $y$ are $k$-dimensional vectors denoting the output and
ground-truth value respectively, and $o_i$ and $y_i$ denote the $i^{th}$
dimension value of $o$ and $y$.  To~train our model, we use \emph{teacher
forcing}, where the target sequence is fed into the decoder directly during
training and used to compute the next output. We~give the full details of
the training approach in the next section.

%% file: simple_network.tikz
\begin{tikzpicture}[x=1.6cm, y=0.25cm, >=stealth]

\definecolor{DKgreen}{RGB}{216, 231, 187}
\definecolor{DKblue}{RGB}{184, 190, 190}
\definecolor{DKred}{RGB}{249, 204, 202}

\fill [color=DKblue] (3.4,-9) rectangle (4.6,7);

\node (in0) [anchor=east] at (1.5, 5) {\small input $i$};
\node (in1) [anchor=east] at (1.5, 3) {\small arg.~1};
\node (in2) [anchor=east] at (1.5, 1) {\small arg.~2};
\node (in3) [anchor=east] at (1.5,-1) {\small arg.~3};
\node (out) [anchor=east] at (1.5,-3) {\small output $i$};

\node (enl1) [xshift=1, draw, rounded corners = 3pt, align=center, minimum height = 2.0cm] at (2,0) {ReLU\\layer\\\scriptsize(256)};

\node (enl2) [xshift=-2, draw, rounded corners = 3pt, align=center, minimum height = 1.6cm] at (3,0) {ReLU\\layer\\\scriptsize(128)};

\node at (2,-8) {\it I/O example encoding};

\node (maxpool) [draw, rounded corners = 3pt, align=center, minimum height = 2cm] at (4,0) {max\\pool};

\node at (4,-8) {\it aggregation};

\node (del1) [xshift=2, draw, rounded corners = 3pt, align=center, minimum height = 2.5cm] at (5,0) {ReLU\\layer\\\scriptsize(512)};

\node (del2) [draw, rounded corners = 3pt, align=center, minimum height = 3cm] at (6,0) {ReLU\\layer\\\scriptsize(5000)};

\node at (6,-8) {\it program decoding};

\node (argmax1) [draw, rounded corners = 3pt, align=center] at (7,6) {arg-\\max};
\node (argmax2) [color=white, align=center] at (7,4) {\makebox{\widthof{arg-}}{}};
\node (argmaxdots) at (7,0) {\vdots};
\node (argmax99) [color=white, align=center] at (7,-4) {\makebox{\widthof{arg-}}{}};
\node (argmax100) [draw, rounded corners = 3pt, align=center] at (7,-6) {arg-\\max};

\node (token1) [anchor=west,xshift=-0.75cm] at (8,6) {token 1};
\node (token100) [anchor=west,xshift=-0.75cm] at (8,-6) {token 100};

\path[draw, ->] (in1.east) -- (enl1);
\path[draw, ->] (in2.east) -- (enl1);
\path[draw, ->] (in3.east) -- (enl1);
\path[draw, ->] (out.east) -- (enl1);
\path[draw, ->] (enl1) -- (enl2);
\path[draw, ->] (enl2) -- (maxpool);
\path[draw, ->] (maxpool) -- (del1);
\path[draw, ->] (del1) -- (del2);
\path[draw, ->] (del2) -- (argmax1.west);
\path[draw, ->] (del2) -- (argmax2.west);
\path[draw, ->] (del2) -- (argmax99.west);
\path[draw, ->] (del2) -- (argmax100.west);
\path[draw, ->] (argmax1) -- (token1);
\path[draw, ->] (argmax100) -- (token100);

\end{tikzpicture}

%% file: paper.bbl
\begin{thebibliography}{10}
\providecommand{\url}[1]{\texttt{#1}}
\providecommand{\urlprefix}{URL }

\bibitem{DBLP:conf/cav/AbateDKKP18}
Abate, A., David, C., Kesseli, P., Kroening, D., Polgreen, E.: Counterexample
  guided inductive synthesis modulo theories. In: {CAV} {(1)}. Lecture Notes in
  Computer Science, vol. 10981, pp. 270--288. Springer (2018)

\bibitem{abolafia2018neural}
Abolafia, D.A., Norouzi, M., Le, Q.V.: Neural program synthesis with priority
  queue training. arXiv preprint arXiv:1801.03526  (2018)

\bibitem{DBLP:conf/cav/AlbarghouthiGK13}
Albarghouthi, A., Gulwani, S., Kincaid, Z.: Recursive program synthesis. In:
  {CAV}. LNCS, vol. 8044, pp. 934--950. Springer (2013)

\bibitem{DBLP:conf/fmcad/AlurBJMRSSSTU13}
Alur, R., Bod{\'{\i}}k, R., Juniwal, G., Martin, M.M.K., Raghothaman, M.,
  Seshia, S.A., Singh, R., Solar{-}Lezama, A., Torlak, E., Udupa, A.:
  Syntax-guided synthesis. In: {FMCAD}. pp. 1--8. {IEEE} (2013)

\bibitem{DBLP:conf/tacas/AlurRU17}
Alur, R., Radhakrishna, A., Udupa, A.: Scaling enumerative program synthesis
  via divide and conquer. In: {TACAS}. LNCS, vol. 10205, pp. 319--336 (2017)

\bibitem{balog2016deepcoder}
Balog, M., Gaunt, A.L., Brockschmidt, M., Nowozin, S., Tarlow, D.: {DeepCoder}:
  Learning to write programs. arXiv preprint arXiv:1611.01989  (2016)

\bibitem{DBLP:conf/cav/BarrettCDHJKRT11}
Barrett, C., Conway, C.L., Deters, M., Hadarean, L., Jovanovic, D., King, T.,
  Reynolds, A., Tinelli, C.: {CVC4}. In: {CAV}. LNCS, vol. 6806, pp. 171--177.
  Springer (2011)

\bibitem{bengio1994learning}
Bengio, Y., Simard, P., Frasconi, P.: Learning long-term dependencies with
  gradient descent is difficult. IEEE Transactions on Neural Networks  5(2),
  157--166 (1994)

\bibitem{DBLP:journals/fac/BradleyM08}
Bradley, A.R., Manna, Z.: Property-directed incremental invariant generation.
  Formal Asp. Comput.  20(4-5),  379--405 (2008)

\bibitem{bunel2018leveraging}
Bunel, R., Hausknecht, M., Devlin, J., Singh, R., Kohli, P.: Leveraging grammar
  and reinforcement learning for neural program synthesis. In: International
  Conference on Learning Representations (2018)

\bibitem{chorowski2015attention}
Chorowski, J.K., Bahdanau, D., Serdyuk, D., Cho, K., Bengio, Y.:
  Attention-based models for speech recognition. In: Advances in neural
  information processing systems. pp. 577--585 (2015)

\bibitem{DBLP:conf/fm/DavidKKL16}
David, C., Kesseli, P., Kroening, D., Lewis, M.: Danger invariants. In: Formal
  Methods ({FM}). LNCS, vol. 9995, pp. 182--198. Springer (2016)

\bibitem{devlin2017robustfill}
Devlin, J., Uesato, J., Bhupatiraju, S., Singh, R., Mohamed, A.R., Kohli, P.:
  Robustfill: Neural program learning under noisy {I/O}. arXiv preprint
  arXiv:1703.07469  (2017)

\bibitem{DBLP:conf/ndss/FengBMDA17}
Feng, Y., Bastani, O., Martins, R., Dillig, I., Anand, S.: Automated synthesis
  of semantic malware signatures using maximum satisfiability. In: {NDSS}. The
  Internet Society (2017)

\bibitem{DBLP:conf/pldi/FengMGDC17}
Feng, Y., Martins, R., Geffen, J.V., Dillig, I., Chaudhuri, S.: Component-based
  synthesis of table consolidation and transformation tasks from examples. In:
  {PLDI}. pp. 422--436. {ACM} (2017)

\bibitem{DBLP:conf/popl/FengM0DR17}
Feng, Y., Martins, R., Wang, Y., Dillig, I., Reps, T.W.: Component-based
  synthesis for complex {APIs}. In: {POPL}. pp. 599--612. {ACM} (2017)

\bibitem{Goodfellow-et-al-2016}
Goodfellow, I., Bengio, Y., Courville, A.: Deep Learning. MIT Press (2016),
  \url{http://www.deeplearningbook.org}

\bibitem{DBLP:conf/pldi/GulwaniJTV11}
Gulwani, S., Jha, S., Tiwari, A., Venkatesan, R.: Synthesis of loop-free
  programs. In: {PLDI}. pp. 62--73. {ACM} (2011)

\bibitem{DBLP:conf/pldi/GulwaniKT11}
Gulwani, S., Korthikanti, V.A., Tiwari, A.: Synthesizing geometry
  constructions. In: {PLDI}. pp. 50--61. {ACM} (2011)

\bibitem{hochreiter1997long}
Hochreiter, S., Schmidhuber, J.: Long short-term memory. Neural computation
  9(8),  1735--1780 (1997)

\bibitem{DBLP:journals/corr/JhaS14a}
Jha, S., Seshia, S.A.: Are there good mistakes? {A} theoretical analysis of
  {CEGIS}. In: {SYNT}. {EPTCS}, vol. 157, pp. 84--99 (2014)

\bibitem{DBLP:journals/corr/KaiserS15}
Kaiser, L., Sutskever, I.: Neural {GPUs} learn algorithms. CoRR  abs/1511.08228
  (2015)

\bibitem{kingma2014adam}
Kingma, D.P., Ba, J.: Adam: A method for stochastic optimization. arXiv
  preprint arXiv:1412.6980  (2014)

\bibitem{mackay2003information}
MacKay, D.J., Mac~Kay, D.J.: Information Theory, Inference and Learning
  Algorithms. Cambridge University Press (2003)

\bibitem{DBLP:conf/tacas/MouraB08}
de~Moura, L.M., Bj{\o}rner, N.: {Z3:} an efficient {SMT} solver. In: {TACAS}.
  Lecture Notes in Computer Science, vol. 4963, pp. 337--340. Springer (2008)

\bibitem{DBLP:conf/pldi/PadhiSM16}
Padhi, S., Sharma, R., Millstein, T.D.: Data-driven precondition inference with
  learned features. In: {PLDI}. pp. 42--56. {ACM} (2016)

\bibitem{parisotto2016neuro}
Parisotto, E., Mohamed, A.r., Singh, R., Li, L., Zhou, D., Kohli, P.:
  Neuro-symbolic program synthesis. arXiv preprint arXiv:1611.01855  (2016)

\bibitem{DBLP:conf/icse/PelegSY18}
Peleg, H., Shoham, S., Yahav, E.: Programming not only by example. In: {ICSE}.
  pp. 1114--1124. {ACM} (2018)

\bibitem{DBLP:conf/pldi/PerelmanGGP14}
Perelman, D., Gulwani, S., Grossman, D., Provost, P.: Test-driven synthesis.
  In: {PLDI}. pp. 408--418. {ACM} (2014)

\bibitem{code2inv}
Si, X., Dai, H., Raghothaman, M., Naik, M., Song, L.: Learning loop invariants
  for program verification. In: {NIPS} (2018)

\bibitem{DBLP:journals/sttt/Solar-Lezama13}
Solar{-}Lezama, A.: Program sketching. {STTT}  15(5-6),  475--495 (2013)

\bibitem{DBLP:conf/asplos/Solar-LezamaTBSS06}
Solar{-}Lezama, A., Tancau, L., Bod{\'{\i}}k, R., Seshia, S.A., Saraswat, V.A.:
  Combinatorial sketching for finite programs. In: {ASPLOS}. pp. 404--415.
  {ACM} (2006)

\bibitem{DBLP:journals/corr/SutskeverVL14}
Sutskever, I., Vinyals, O., Le, Q.V.: Sequence to sequence learning with neural
  networks. In: {NIPS}. pp. 3104--3112 (2014)

\bibitem{dataprep}
Yu, L., Wang, S., Lai, K.K.: An integrated data preparation scheme for neural
  network data analysis. IEEE Transactions on Knowledge and Data Engineering
  18,  217-- 230 (03 2006)

\end{thebibliography}
